\documentclass[prd,twocolumn,showpacs,preprintnumbers,floats,floatfix,nofootinbib]{revtex4-1}
\usepackage{graphicx}
\usepackage{dcolumn}
\usepackage{latexsym}
\usepackage{amsmath}

\begin{document}

\title{Puncture black hole initial data: a single domain Galerkin-Collocation method for trumpet and wormhole data sets}
\author{P. C. M. Clemente and H. P. de Oliveira}
\email{hp.deoliveira@pq.cnpq.br}
\affiliation{
Departamento de F\'{\i}sica Te\'orica - Instituto de F\'{\i}sica
A. D. Tavares, Universidade do Estado do Rio de Janeiro \\ 
R. S\~ao Francisco Xavier, 524. Rio de Janeiro, RJ, 20550-013, Brazil}


\date{\today}

\begin{abstract}
We present a single domain Galerkin-Collocation method to calculate puncture initial data sets for single and binary, either in the trumpet or wormhole geometries. The combination of aspects belonging to the Galerkin and the Collocation methods together with the adoption of spherical coordinates in all cases show to be very effective. We have proposed a unified expression for the conformal factor to describe trumpet and spinning black holes. In particular, for the spinning trumpet black holes, we have exhibited the deformation of the limit surface due to the spin from a sphere to an oblate spheroid. We have also revisited the energy content in the trumpet and wormhole puncture data sets. The algorithm can be extended to describe binary black holes. 
\end{abstract}


\maketitle

\section{Introduction}%

The precise characterization of the gravitational and matter fields on some spatial hypersurface constitutes the initial data problem in numerical relativity \cite{cook}. In this instance, it is possible to identify if there exist interacting black holes and neutrons stars together or not with any other distribution of matter, placing an ideal set up to simulate astrophysical situations in which the high gravitational field plays a central role. In parallel to the decades-long effort to directly detect gravitational radiation which has been accomplished recently \cite{LIGO_gws}, there has also been an endeavor to predict gravitational wave signals from compact binaries using numerical simulations. These simulations \cite{pretorius,campanelli,baker} start with initial data in general containing binary of black holes 

In more precise terms, the initial data problem in General Relativity consists in specifying the spatial metric and extrinsic curvature, $\gamma_{ij}$ and $K_{ij}$, respectively, in a given spatial hypersurface. These quantities must satisfy the constraint equations, namely the Hamiltonian and momentum constraint equations, that arises from the Cauchy formulation of the field equations \cite{adm}. The most important strategy for solving the constrained equations is to introduce a conformal transformation of the spatial metric to a known background metric, $\bar{\gamma}_{ij}$, and a similar transformation involving the extrinsic curvature \cite{york_1979}. Then,

\begin{eqnarray}
\gamma_{ij}&=&\Psi^4 \bar{\gamma}_{ij} \label{eq1}\\
\nonumber \\
A_{ij}&=&\Psi^{-2} \bar{A}_{ij}, \label{eq2}
\end{eqnarray}

\noindent where $\Psi$ is the confomal factor and $A_{ij}$ is the traceless part of the extrinsic curvature such that,

\begin{equation}
K_{ij} = A_{ij} + \frac{1}{3} \gamma_{ij} K, \label{eq3}
\end{equation}

\noindent with $K$ being the trace of $K_{ij}$. In this formulation, the set of functions $(\Psi,\bar{\gamma}_{ij},\bar{A}_{ij},K)$ specified in the initial hypersurface characterizes the initial data. These quantities are not fixed by the constraint equations but must them. We assume here the Bowen-York \cite{bowen_york} scheme, that is, the additional requirements of conformal flatness, maximal slicing, $K=0$, and vacuum, yielding the decoupling of the Hamiltonian and momentum constraints which, respectively, become,

\begin{eqnarray}
\bar{\nabla}^2 \Psi + \frac{1}{8} \Psi^{-7}  \bar{A}^{ij} \bar{A}_{ij} = 0 \label{eq4} \\
\nonumber \\
\bar{D}_i \bar{A}^{ij} = 0, \label{eq5}
\end{eqnarray}

\noindent where $\bar{D}_i = \bar{\gamma}_{ij} \nabla^j$ is the covariant derivative associated with the flat background metric $\bar{\gamma}_{ij}$ and  $\bar{\nabla}^2$ is the flat-space Laplacian operator. Remarkably, Eq. (\ref{eq4}) can be solved analytically to describe boosted and spinning black holes denoted by $\bar{A}^{ij}_\mathbf{P}$ and $\bar{A}^{ij}_\mathbf{S}$, whose corresponding expressions are,

\begin{eqnarray}
\bar{A}^{ij}_{\mathbf{P}} &=& \frac{3}{2 r^2} \left[2P^{(i} n^{j)}-(\eta^{ij}-n^i n^j) \mathbf{n}.\mathbf{P}\right] \label{eq6}\\
\nonumber \\
\bar{A}^{ij}_{\mathbf{S}} &=& \frac{6}{r^3} n^{(i}\epsilon^{j)}_{mp}J^m n^p, \label{eq7}
\end{eqnarray}

\noindent where $\mathbf{P}$ and $\mathbf{J}$ are, respectively, the ADM linear and angular momenta carried by the black hole \cite{baumgarte_shapiro}. The quantity $n^k=x^k/r$ is the normal vector pointing away from the black hole located at $r=0$. Due to the linearity of the momentum constraint, we can construct spacetimes containing a boosted-spinning black hole or multiple black holes by superposing several conformal extrinsic curvature given by Eqs. (\ref{eq5}) and (\ref{eq6}).






In general, the Hamiltonian constraint (\ref{eq4}) is solved numerically for the conformal factor after specifying the extrinsic curvature $\bar{A}_{ij}$. To guarantee that there are black holes in the initial hypersurface it is necessary to satisfy appropriate boundary conditions which are dictated by the excision or puncture methods. We are going to focus here on the puncture method that consists \cite{brandt_brugmann} in decomposing the conformal factor into two pieces: the background component containing the black holes singularities and usually given analytically, and the regular component which is obtained by solving the Hamiltonian constraint numerically. Accordingly, we have,

\begin{equation}
\Psi = \Psi_0 + u. \label{eq8}
\end{equation}

\noindent Considering a single black hole, $\Psi_0$ is taken as the Schwarzschild black hole in its wormhole representation,  or equivalently on a slice of constant Schwarzschild time. It means that,

\begin{equation}
\Psi_0=1+\frac{m_0}{2r}, \label{eq9}
\end{equation}

\noindent where $r=0$ locates the puncture and $m_0$ is a free parameter. It can be verified that the above expression is the solution of the Hamiltonian constraint for $\bar{A}_{ij}=0$ and $u=0$, and in this situation the parameter $m_0$ is the ADM mass. The substitution of Eqs. (\ref{eq8}) and (\ref{eq9}) into the Hamiltonian constraint (\ref{eq4}) results in an elliptic equation for the regular component $u$. We can construct initial data with multiple black holes by a direct generalization of the background conformal factor to $\Psi_0 = 1 + \sum_k m_k/2r_k$, where each puncture $m_k$ located at $r_k=0$. Of particular interest is the case of binary black holes for which most of the initial data used in the simulations adopt the puncture method \cite{campanelli,brugmann_04,baumg_2000,diener_06,baker_06,meter_06,bode_09}.


There is another representation of the Schwarzschild black hole based on spatial slices that terminate at non-zero areal radius known as the trumpet representation. The interest in constructing trumpet initial data has increased after the advent of the moving puncture method \cite{campanelli,baker}. It has been shown that the Schwarzschild wormhole puncture data evolves in such a way the numerical slices tend a spatial slice with finite areal radius or trumpets \cite{hannan_mov_punct,hannan_mov_punct2,brown,hannan_mov_punct3}. Therefore, it is motivating to construct initial trumpet data for single and binary black holes endowed with spin and linear momentum. In this direction, we mention the derivation of the analytical solutions for maximally sliced and 1+log trumpet Schwarzschild black holes in Refs. \cite{baumg_nac,denninson_baumg}, respectively. The initial data for spinning boosted, single and binary trumpets were studied by Hannan et al. \cite{hanann_id_trumpet}, Immerman and Baumgarte \cite{immer_baumg} for the maximally sliced case. More recently, Dietrich and Brugman \cite{dietrich_brugmann} constructed 1+log sliced initial data for single and binary systems.


We present here a single domain algorithm based on Galerkin-Collocation spectral method \cite{deol_rod_bondi,deol_rod_RT,deol_rod_idata2} to obtain wormhole and trumpet initial data sets. The algorithm is distinct from other spectral codes \cite{pfeiffer_CPC,ansorg_1,ansorg_07,ossokine}, but nonetheless very efficient and simple. We believe that this task is valuable in its own right. The selection of the radial and angular basis functions is of crucial importance; we have the spherical harmonics as the most natural basis functions for the angular domain, whereas the radial basis functions are expressed as appropriate linear combinations of the Chebyshev polynomials to satisfy the boundary conditions. The algorithm is well suited to describe spinning and boosted single black hole, a wormhole or a trumpet binary system. 

The paper is divided as follows. After the Introduction in Section 1, we have focused on presenting the basic equations for constructing trumpet initial data sets. We have used the maximal sliced analytical solution of Naculich and Baumgarte \cite{baumg_nac} to establish a convenient expression for the conformal factor describing single or binary trumpets. The numerical scheme is detailed in Section 3. We have presented the numerical tests and discussed some cases of interest in Section 4. In particular, we highlighted the proposed unified description of single trumpet spinning and trumped black hole. For a single spinning black hole, we have shown the influence of the spin in altering the minimal surface from a sphere to an oblate spheroid. We have also considered wormhole and trumpet binaries to illustrate the feasibility of the algorithm in more general cases. Finally, in Section 5 we have concluded and traced some directions of the present investigation.

\section{Trumpet and wormhole puncture data sets} %

The starting point to construct maximal sliced puncture trumpet initial data is to establish the trumpet slicing of the Schwarzschild spacetime. Baumgarte and Naculich \cite{baumg_nac} have derived the corresponding exact conformal factor in function of the areal radius $R = r \Psi_0^2$ (cf. Appendix A). With the exact solution, they have shown following asymptotic behavior, 

\begin{eqnarray}
\Psi_0 &=& \left(\frac{3 m_0}{2 r}\right)^{1/2},\;\; r \rightarrow 0 \label{eq10}\\
\nonumber \\
\Psi_0 &=& 1 + \frac{m_0}{2r},\;\; r \rightarrow \infty, \label{eq11}
\end{eqnarray}

\noindent where $m_0$ is the Schwarzschild mass. The corresponding expression for the traceless part of the extrinsic curvature is,

\begin{equation}
\bar{A}_0^{ij} = \frac{3 \sqrt{3}m_0^2}{4 r^3} (\bar{\gamma}^{ij} - 3 n^i n^j). \label{eq12}
\end{equation}

\noindent In the case of wormhole data we have $\bar{A}_0^{ij}=0$. With the above expression it can be shown that the momentum constraint $\bar{D}_i \bar{A}_0^{ij} = 0$ is satisfied along with the validity of the Hamiltonian constraint, 

\begin{equation}
\bar{\nabla}^2 \Psi_0 + \frac{1}{8}\Psi_0^{-7} \bar{A}^{ij}_0 \bar{A}^0_{ij} = 0, \label{eq13}
\end{equation} 

\noindent where,

\begin{equation}
\bar{A}^{ij}_0 \bar{A}^0_{ij} = \frac{81 m_0^4}{8r^6}. \label{eq14}
\end{equation}

For the trumpet initial data sets, we propose the following puncture-like expression for the conformal factor,

\begin{equation}
\Psi = \Psi_0(1+u), \label{eq15}
\end{equation}

\noindent where $\Psi_0$ is trumpet Schwarzschild solution. Introducing the new conformal factor into the Hamiltonian constraint (\ref{eq4}), we have,

\begin{eqnarray}
\bar{\nabla}^2 u + \frac{2\bar{D}_i\Psi_0\bar{D}^i u}{\Psi_0} + \frac{\bar{A}^{ij}\bar{A}_{ij}}{8 \Psi_0^8(1+u)^7} - \frac{(1+u)}{8 \Psi_0^8}\bar{A}_0^{ij}\bar{A}^0_{ij} = 0,\label{eq16} \nonumber \\
\end{eqnarray}

\noindent where the total traceless part of the extrinsic curvature is given by, 

\begin{equation}
\bar{A}^{ij} = \bar{A}^{ij}_0 + \bar{A}^{ij}_{\mathbf{P}} + \bar{A}^{ij}_{\mathbf{S}}, \label{eq17}
\end{equation}

\noindent due to the linearity of the momentum constraint equation. In the case of the wormhole data sets the conformal factor is expressed by (\ref{eq8}) and the Hamiltonian equation becomes, 

\begin{eqnarray}
\bar{\nabla}^2 u + \frac{1}{8}(\Psi_0+u)^{-7}\bar{A}^{ij}\bar{A}_{ij} = 0, \label{eq18}
\end{eqnarray}

\noindent with $\bar{A}^{ij}_0=0$ and $\Psi_0$ given by Eq. (\ref{eq9}).

The main reason of not adopting the usual decomposition for the conformal factor (Eq. (\ref{eq8})) for trumpet black hole data sets is to provide an unified framework for describing spinning and boosted black holes with regular functions $u$. For instance, for a single trumpet spinning black hole in which $\Psi=\Psi_0+u$, it can be shown that \cite{immer_baumg,hanann_id_trumpet} $u \sim \mathcal{O}(r^{-1/2})$ near $r=0$, and for a single boosted black hole $u \sim \mathcal{O}(r)$. On the other hand, by considering the new decomposition (\ref{eq12}), we have followed the analysis of Immerman and Baumgarte \cite{immer_baumg} of the behavior of $u$ near the puncture at $r=0$ for a boosted ($u_P$) and a spinning black hole ($u_S$) in the axisymmetric case. Assuming that $u \ll 1$, the corresponding Hamiltonian constraints are approximated by,

{\small
\begin{eqnarray}
&&\bar{\nabla}^2 u_P - \frac{1}{r}\frac{\partial u_P}{\partial r} \approx \frac{\sqrt{3}P\cos\theta}{3m_0^2 r} + 2\frac{u_P}{r^2} \label{eq19}\\
\nonumber \\
&&\bar{\nabla}^2 u_S - \frac{1}{r}\frac{\partial u_S}{\partial r} \approx -\frac{4 J^2 \sin^2\theta}{9 m_0^4 r^2} + \left(1+\frac{28J^2\sin^2\theta}{9m_0^4}\right)\frac{u_S}{r^2}. \label{eq20} \nonumber \\
\end{eqnarray}
}

\noindent From these equations one can show that near the origin,

\begin{equation}
u_P \sim \mathcal{O}(r),\;\mathrm{and}\;u_S \sim \mathcal{O}(1). \label{eq21}
\end{equation}

\noindent The above behaviors near the origin can be dealt numerically without difficulties. 

To guarantee that the spacetime is asymptotically flat, the function $u$ must satisfy the following asymptotic condition,

\begin{equation}
u = \frac{\delta m}{r} + \mathcal{O}(r^{-2}), \label{eq22}
\end{equation}

\noindent where $\delta=\delta(\theta,\phi)$ in general after adopting the spherical coordinates. As indicated in the sequence, the function $\delta m$ is the contribution due to angular and linear momenta to the ADM mass which is calculated from,

\begin{equation}
M_{ADM} = - \frac{1}{2\pi}\,\lim_{r \rightarrow \infty} \int_{\Omega}\,r^2 \Psi_{,r} d \Omega. \label{eq23}
\end{equation}

\noindent Assuming the conformal factor either expressed by Eq. (\ref{eq8}) or Eq. (\ref{eq15}), and taking into account the behavior of $u$ and $\Psi_0$ for $r \rightarrow \infty$, we obtain,

\begin{equation}
M_{ADM} = m_0 + \frac{1}{2\pi}\,\int_0^{2\pi}\int_0^\pi\,\delta m(\theta,\phi) \sin \theta\, d\theta d\phi. \label{eq24}
\end{equation}

\noindent According to the numerical scheme of next Section, we can read off an analytical expression for $\delta m(\theta,\phi)$, and the ADM mass is calculated straightforwardly. In the case of multiple black holes, we have to replace $m_0 \rightarrow \sum m_i$ in the above expression.

\section{The Galerkin-Collocation algorithm}%

We present here the Galerkin-Collocation scheme to solve the Hamiltonian constraint (\ref{eq16}) or (\ref{eq18}) for trumpet and wormhole data sets. The centerpiece of the numerical treatment is the spectral approximation of the function $u(r,\theta,\phi)$ given by, 

\begin{equation}
u_a(r,\theta,\phi) = \sum^{N_x,N_y}_{k,l = 0}\sum^{l}_{m=-l}\,c_{klm} ~\chi_{k}(r) Y_{lm}(\theta,\phi). \label{eq25}
\end{equation}

\noindent Here $c_{klm}$ represents the unknown coefficients or modes, $N_x$ and $N_y$ are, respectively, the radial and angular truncation orders that limit the number of terms in the above expansion. The angular patch has the spherical harmonics, $Y_{lm}(\theta,\phi)$, as the basis functions. The choice of spherical coordinates together with the adoption of spherical harmonics basis functions are quite natural, and as we are going to show, are computationally very efficient.  Concerning the radial basis functions, ${\chi_k(r)}$, we have followed the prescription of the Galerkin method in which each basis function satisfies the boundary conditions. Usually, this is done by establishing an appropriate combination of Chebyshev polynomials. Near $r=0$, we have,

\begin{eqnarray}
\chi_k(r) \sim \mathcal{O}(r),\; \mathrm{and}\; \chi_k(r) \sim \mathcal{O}(1), \label{eq26}
\end{eqnarray}

\noindent according with the boundary conditions (\ref{eq21}). The asymptotic behavior of each basis function is,

\begin{equation}
\chi_k(r) \sim \mathcal{O}(r^{-1}). \label{eq27}
\end{equation}

\noindent To satisfy these boundary conditions, we define each radial basis function as,

\begin{eqnarray}
\chi_k(r) = \frac{1}{2}(TL_{k+2}(r)-TL_{k}(r)), \label{eq28}\\
\nonumber \\
\chi_k(r) = \frac{1}{2}(TL_{k+1}(r)-TL_{k}(r)), \label{eq29}
\end{eqnarray}

\noindent for boosted and spinning black holes, respectively. For the wormhole case the basis function is given by expression (\ref{eq29}). Here  $TL_k(r)$ represents the rational Chebyshev polynomials defined by,

\begin{eqnarray}
TL_k(r) = T_k\left(x = \frac{r - L_0}{r+L_0}\right) \label{eq30}
\end{eqnarray}  

\noindent where $T_k(x)$ is the Chebyshev polynomial of $k$th order and $L_0$ is the map parameter that connects $-1 \leq x <1$ to $0 \leq r < \infty$ through the algebraic map \cite{boyd} $r=L_0 (1+x)/(1-x)$.

The spherical harmonics are complex functions implying that the coefficients $c_{klm}$ must be complex but satisfying some symmetry conditions to guarantee that the conformal factor be a real function. The symmetry conditions are,

\begin{equation}
c^*_{kl-m}=(-1)^{-m}\,c_{klm}, \label{eq31}
\end{equation}

\noindent due to the symmetry relation of the spherical harmonics $Y^*_{l-m}(\theta,\phi)=(-1)^{-m}Y_{lm}(\theta,\phi)$. Consequently, the number of independent modes $\left(N_x+1\right)\left(N_y+1\right)^2$.

We now establish the residual equation associated with the Hamiltonian constraint by substituting the spectral approximation (\ref{eq21}) into the Hamiltonian constraint (\ref{eq16}) (or (\ref{eq18})). In addition, we have taken into account the differential equation for the spherical harmonics to get rid of the derivatives with respect to $\theta$ and $\phi$. After a straighforward calculation, we have arrived to the following expression,

\begin{widetext}
{\small{
\begin{eqnarray}
\mathrm{Res}(r,\theta,\phi)&=&\sum_{k,n,p}\,c_{knp}\left[\frac{1}{r^2}\frac{\partial}{\partial r}\left(r^2 \frac{\partial \chi_k}{\partial r}\right) - 
\frac{n(n+1)}{r^2}\chi_k \right] Y_{np}(\theta,\phi) + \frac{2}{\Psi_0}\frac{\partial \Psi_0}{\partial R}\frac{\partial R}{\partial r}\frac{\partial u_a}{\partial r} -\frac{(1+u_a(r,\theta,\phi))}{8\Psi_0^8}(\bar{A}^{ij}\bar{A}_{ij})_0 + \nonumber \\
\nonumber \\
&&+ \frac{(1+u_a(r,\theta,\phi))^{-7}}{8\Psi_0^8}\bar{A}^{ij}\bar{A}_{ij}. \label{eq32}
\end{eqnarray}}
}
\end{widetext}

\noindent In the case of binary systems with trumpet punctures, it is necessary to modify the second term on the RHS to include the angular dependence that appears in the background solution $\Psi_0$.



\begin{figure}
\includegraphics[height=5.cm,width=6.cm]{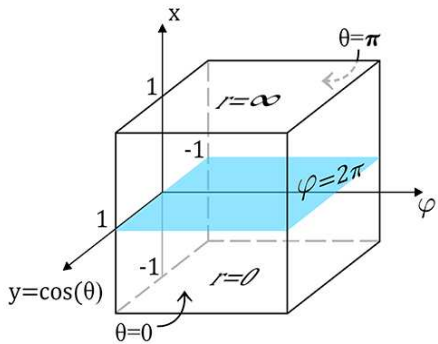}
\caption{The three-dimensional spatial domain viewed as a cube described by the coordinates $-1 \leq x < 1$, $-1 < y < 1$ ($y=\cos \theta$) that correspond to $0 \leq r < \infty$ and  $0 \leq \theta \leq \pi$, while the azimuthal angle $\phi$ is maintained.}
\end{figure}

The next and final step is to describe the procedure to obtain de coefficients $c_{klm}$. From the method of weighted residuals \cite{finlayson}, these coefficients are evaluated with the condition of forcing the residual equation to be zero in an average sense. It means that,

\begin{eqnarray}
&& \left<\mathrm{Res},R_j(r)S_{lm}(\theta,\phi)\right> = \nonumber \\
\nonumber \\
&& = \int_{\mathcal{D}}\,\mathrm{Res} R^*_j(r) S^{*}_{lm}(\theta,\phi)\,w_r w_\theta w_\phi \,dr d\Omega=0,   \label{eq33}
\end{eqnarray}

\noindent where the functions $R_j(r)$ and $S_{lm}(\theta,\phi)$ are called the test functions while $w_r, w_\theta$ and $w_\phi$ are the corresponding weights. We have chosen the radial test function as prescribed by the Collocation method,

\begin{equation}
R_j(r) = \delta(r-r_j), \label{eq34}
\end{equation}

\noindent which is the delta of Dirac function; $r_j$ represents the radial collocation points and $w_r=1$. Following the Galerkin method we identify the angular test function $S_{lm}(\theta,\phi)$ as the spherical harmonics, and consequently $w_\theta=w_\phi=1$. Therefore Eq. (\ref{eq33}) becomes,

\begin{equation}
\left<\mathrm{Res}(r,\theta,\phi),Y_{lm}(\theta,\phi)\right>_{r=r_j}=0, \label{eq35}
\end{equation}

\noindent where $j=0,1,..,N_x$, $l=0,1,..,N_y$ and $m=0,1,..,l$. The $N_x+1$ radial collocation points are,

\begin{eqnarray}
r_j = \frac{L_0 (1+\tilde{x}_j)}{1-\tilde{x}_j}. \label{eq36}
\end{eqnarray}

\noindent with the Chebyshev-Gauss collocation points $\tilde{x}_j$ in the computational domain, 

\begin{eqnarray}
\tilde{x}_j=\cos\left[\frac{(2 j+1) \pi}{2 N_x+2}\right],\; j=0,1,..,N_x.\label{eq37}
\end{eqnarray}

\noindent We have excluded the point at infinity ($\tilde{x}=1$) since the residual equation (\ref{eq32}) is identically satisfied asymptotically due to the choice of the radial basis functions. Noticed that the origin is also excluded. In Fig. 1 we show schematically the spatial domain spanned by the new coordinates $(\tilde{x},y=\cos \theta,\phi)$.  

We are in conditions to present schematically the set of equations resulting from the relations (\ref{eq35}). The integration on the angular domain takes into account the orthogonality of the spherical harmonics in the first three terms of the residual equation (\ref{eq32}) whose result is,

{\small
\begin{widetext}
\begin{eqnarray}
\left<\mathrm{Res},Y_{lm}(\theta,\phi)\right>_{r_j} = \sum_{k}\,c_{klm}\left[\frac{1}{r^2}\frac{\partial}{\partial r}\left(r^2 \frac{\partial \chi_k}{\partial r}\right) - \frac{l(l+1)\chi_k}{r^2} \right]_{r_j} + \left(\frac{2}{\Psi_0}\frac{\partial \Psi_0}{\partial R}\frac{\partial R}{\partial r}\right)_{r_j}\sum_k\,c_{klm}\left(\frac{\partial \chi_k}{\partial r}\right)_{r_j} \nonumber \\
\nonumber\\
-\left(\frac{(\bar{A}^{ij}\bar{A}_{ij})_0}{8\Psi_0^8}\right)_{r_j}(2\sqrt{\pi} \delta_{0l}\delta_{0m}+\sum_k\,c_{klm}\chi_k(r_j)) + \left<\frac{\left(\bar{A}^{ij}\bar{A}_{ij}\right)}{8\Psi_0^8(1+u_a)^7},Y_{lm}(\theta,\phi)\right>_{r_j}=0, \nonumber \\ \label{eq38}
\end{eqnarray}
\end{widetext}
}

\noindent with $j=0,1,..,N_x$, $l=0,1,..,N_y$ and $m=-l,..,l$. The last term is calculated using quadrature formulae as indicated below,

\begin{eqnarray}
\left<(..),Y_{lm}(\theta,\phi)\right>_{r_j} \approx \sum_{k,n=0}^{N_1,N_2}\,(..) Y^*_{lm}(\theta_k,\phi_n) v^{\theta}_k v^{\phi}_n, \label{eq39}
\end{eqnarray}

\noindent where $(\theta_k,\phi_n)$, $k=0,1,..,N_1$, $n=0,1,..,N_2$ are the quadrature collocation points, and $v^{\theta}_k v^{\phi}_n$ are the corresponding weights \cite{fornberg}. To achieve better accuracy we have set $N_1=N_2=2 N_y+1$, but this is not mandatory since it is possible to use simply $N_1=N_2=N_y$. In summary, we have to solve the set of $(N_x+1) (N_y+1)^2$ nonlinear algebraic equations indicated by expression (\ref{eq38}) for an equal number of coefficients $c_{klm}$. For that aim the Newton-Raphson algorithm was employed.

\section{Applications}%

\subsection{Single spinning and boosted black holes}

We begin by considering a single spinning or a boosted black hole located at the origin $r=0$. In each case the angular and linear momenta lie on the z-axis, that is $\mathbf{J} = (0,0,J_0)$ and $\mathbf{P} = (0,0,P_0)$. The quantities $A_{ij}A^{ij}$ corresponding to spinning and boosted black holes are given by, 

{\small 
\begin{eqnarray}
\bar{A}_{ij}\bar{A}^{ij} &=& \frac{18 J_0^2}{r^6}\sin^2\theta + \frac{81 m_0^4}{8 r^6} \label{eq40}\\
\nonumber \\
\bar{A}_{ij}\bar{A}^{ij} &=& \frac{9 P_0^2}{2r^4}(1+2\cos^2\theta)+\frac{81 m_0^4}{8 r^6}-\frac{27\sqrt{3}m_0^2P_0}{2r^5}\cos\theta. \nonumber \label{eq41}
\\
\end{eqnarray}}

\noindent The resulting Hamiltonian constraint in each case is axisymmetric due to the absence of any dependence of the polar angle $\phi$. Thus, in the spectral approximation of the function $u(r,\theta)$ (cf. Eq. (\ref{eq25})) the spherical harmonics are replaced by Legendre polynomials as the angular basis functions.


We have adopted the convergence of the ADM mass evaluated according to Eq. (\ref{eq24}) as the main numerical test. From the spectral approximation (\ref{eq25}) we can obtain $\delta m(\theta)$ after $-\lim_{r \rightarrow \infty}\,r^2 \partial u_a(r,\theta)/\partial r$ without approximating the infinity to some finite radius $r_{\mathrm{max}}$. We have established the convergence of the ADM mass by calculating the difference of the ADM mass corresponding to approximate solutions with fixed $N_y=12$ and varying $N_x=5,10,15,..$ such that $\delta M(N_x) = |M_{ADM}(N_x+5)-M_{ADM}(N_x)|$. As reported previously \cite{deol_rod_idata2} the value of the map parameter can improve the convergence of $\delta M$. Figs. 2 and 3 show the convergence tests for spinning and boosted black holes, respectively, where in both cases $m_0=1.0$; the spin parameter is $J_0=0.5 m_0^2$ while the boost is $P_0=1.0m_0$. In Fig. 2 the results are displayed for $L_0=2.0$ and $L_0=0.2$ for the trumpet data sets to illustrate the role of $L_0$ in the convergence rate. Noticed the improvement of the convergence rate is achieved when $L_0=0.2$. For the spinning wormhole, the best map parameter is $L_0=0.5$, and the convergence is better than in the trumpet case. Fig. 3 show the convergence of the ADM mass for trumpet and wormhole boosted black holes with their respective best map parameters, $L_0=0.1$ and $L_0=2.0$.

\begin{figure}
\includegraphics[height=5.cm,width=6.cm]{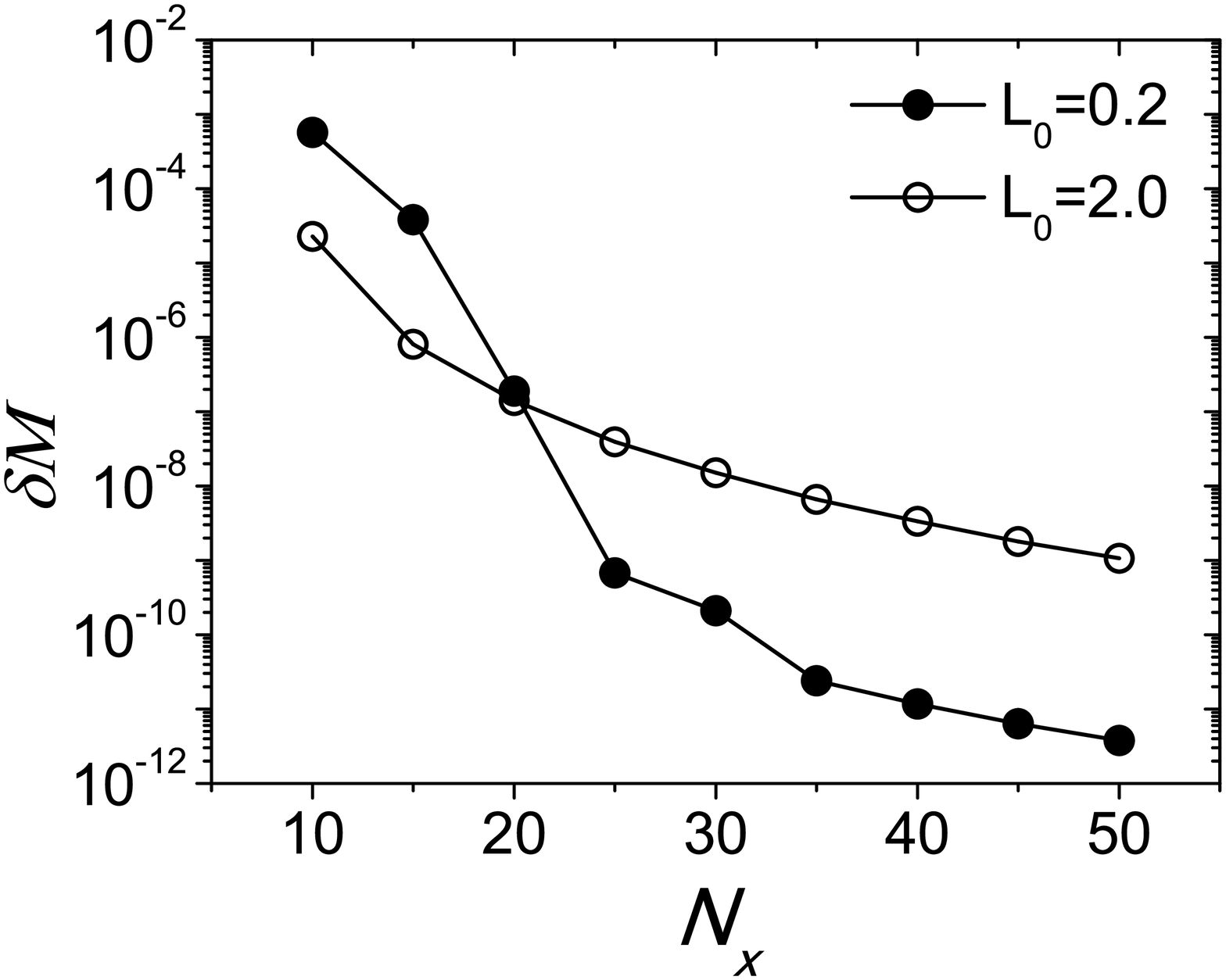}
\includegraphics[height=5.cm,width=6.cm]{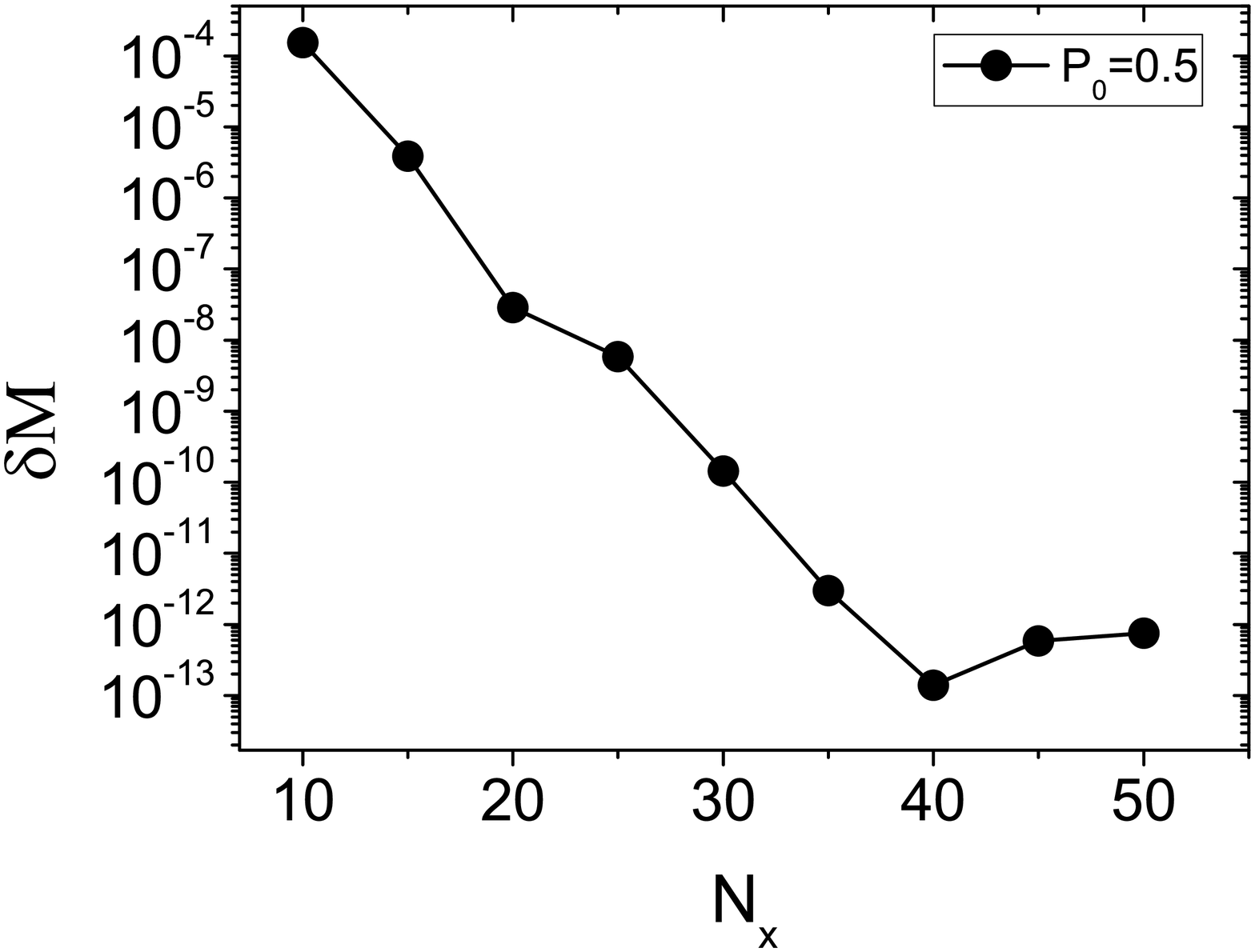}
\caption{Convergence of the ADM mass for trumpet and wormhole spinning punctures (upper and lower graphs, respectively). Here $J_0=0.5 m_0^2$ and $m_0=1.0$. For the trumpet data, we have included two convergence tests corresponding to $L_0=0.2$ and $L_0=2.0$ to make clear the influence of the map parameter. The exponential convergence of the ADM mass is more evident for $L_0=0.2$ than for $L_0=2.0$ the convergence is algebraic. For the wormhole spinning puncture the convergence clearly exponential.}
\end{figure}

\begin{figure}
\includegraphics[height=5.cm,width=6cm]{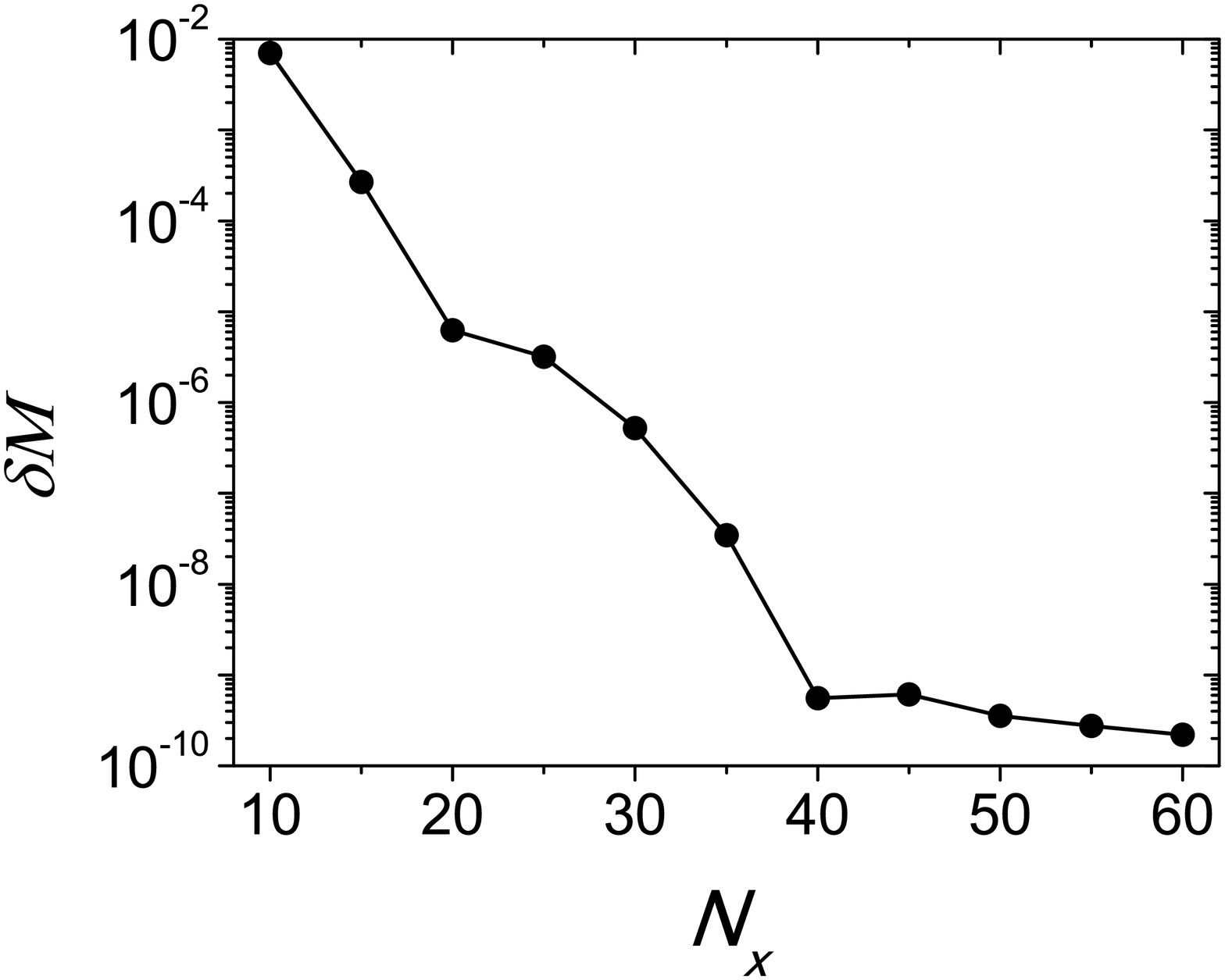}
\includegraphics[height=5.cm,width=6cm]{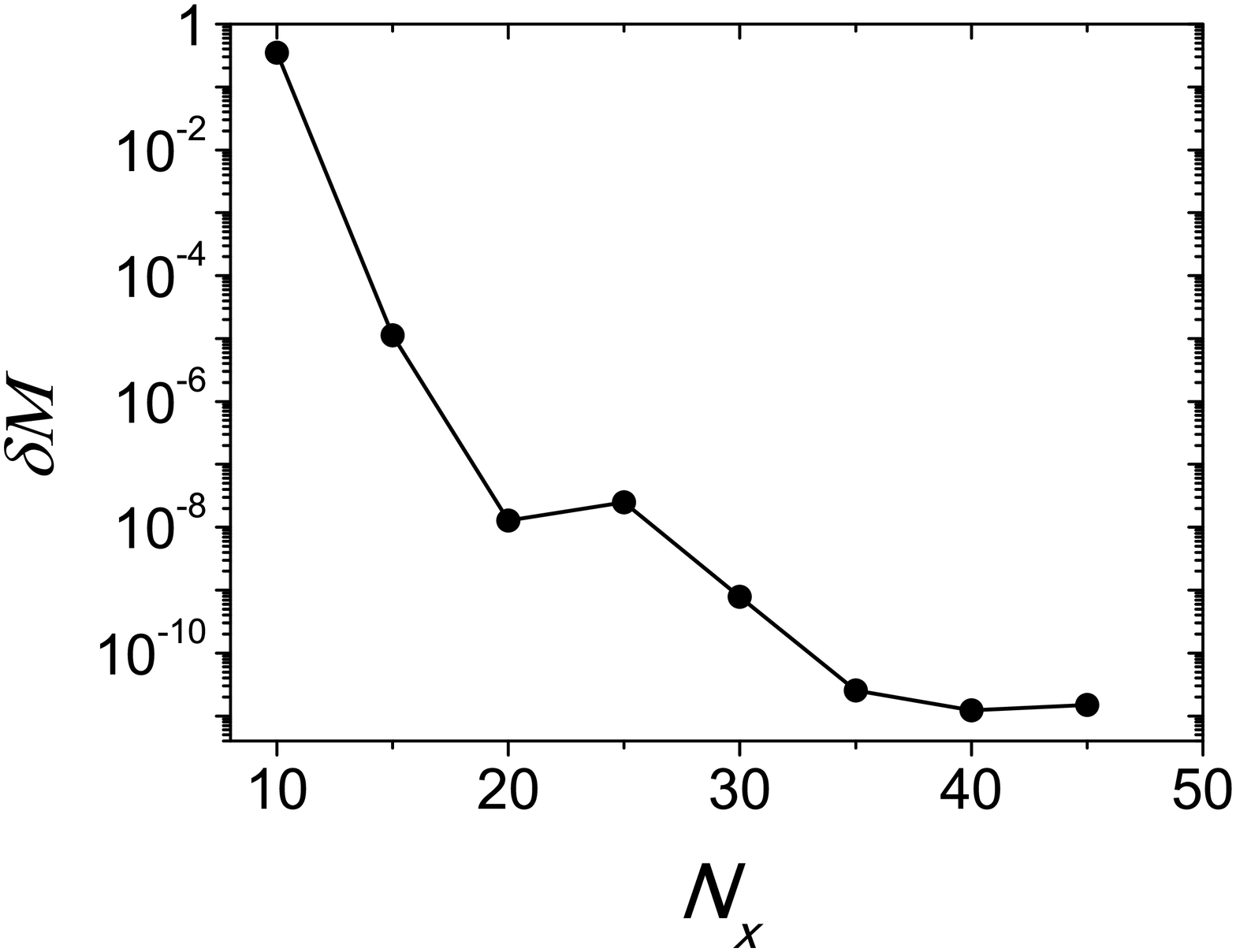}
\caption{Convergence of the ADM mass for trumpet and wormhole boosted punctures (upper and lower graphs, respectively). Here $P_0=1.0 m_0$ and $m_0=1.0$ and the map parameter are $L_0=0.1$ and $L_0=1.0$, respectively. The exponential convergence is achieved in both cases.}
\end{figure}

Spinning trumped black holes alter the geometry of the minimal surface characterized by $r=0$ from spherical to an oblate spheroid. It will be instructive to quantity this change by evaluating the eccentricity of the spheroid in function of the spin parameter $J_0$. The eccentricity of the minimal surface is defined by,

\begin{equation}
\epsilon = \sqrt{1 - \frac{R^2_{\mathrm{min}}(J_0,\theta=0)}{R^2_{\mathrm{min}}(J_0,\theta=\pi/2)}}, \label{eq42}
\end{equation}

\noindent where $R_{\mathrm{min}}(J_0,\theta) = \lim_{r \rightarrow 0}r \Psi_0^2(1+u)^2$. We have expressed the eccentricity in function of $J_0/m_0^2$ and $J_0/M_{ADM}^2$ in Fig. 4. Notice that the eccentricity tends to a limit value of $\epsilon \approx 0.439$. We have included an inset plot with the eccentricity calculated from the approximate solution due to Immerman and Baumgarte \cite{immer_baumg} (continuous line) valid for small $J_0$ and the corresponding numerical eccentricities (circles). As expected, it becomes evident the disagreement between both results as the spin increases.

\begin{figure}
\includegraphics[height=5.cm,width=7.0cm]{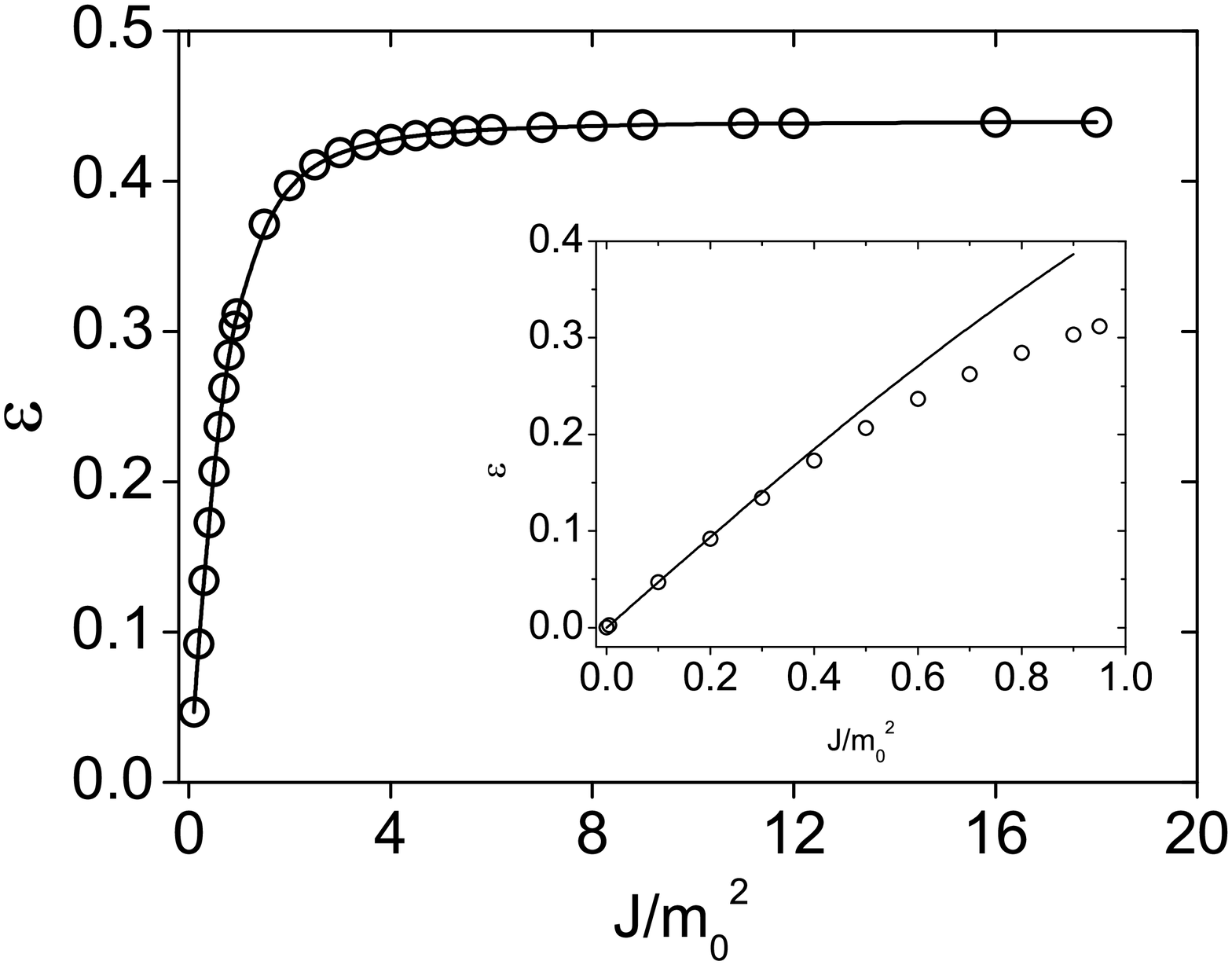}
\includegraphics[height=5.cm,width=7.0cm]{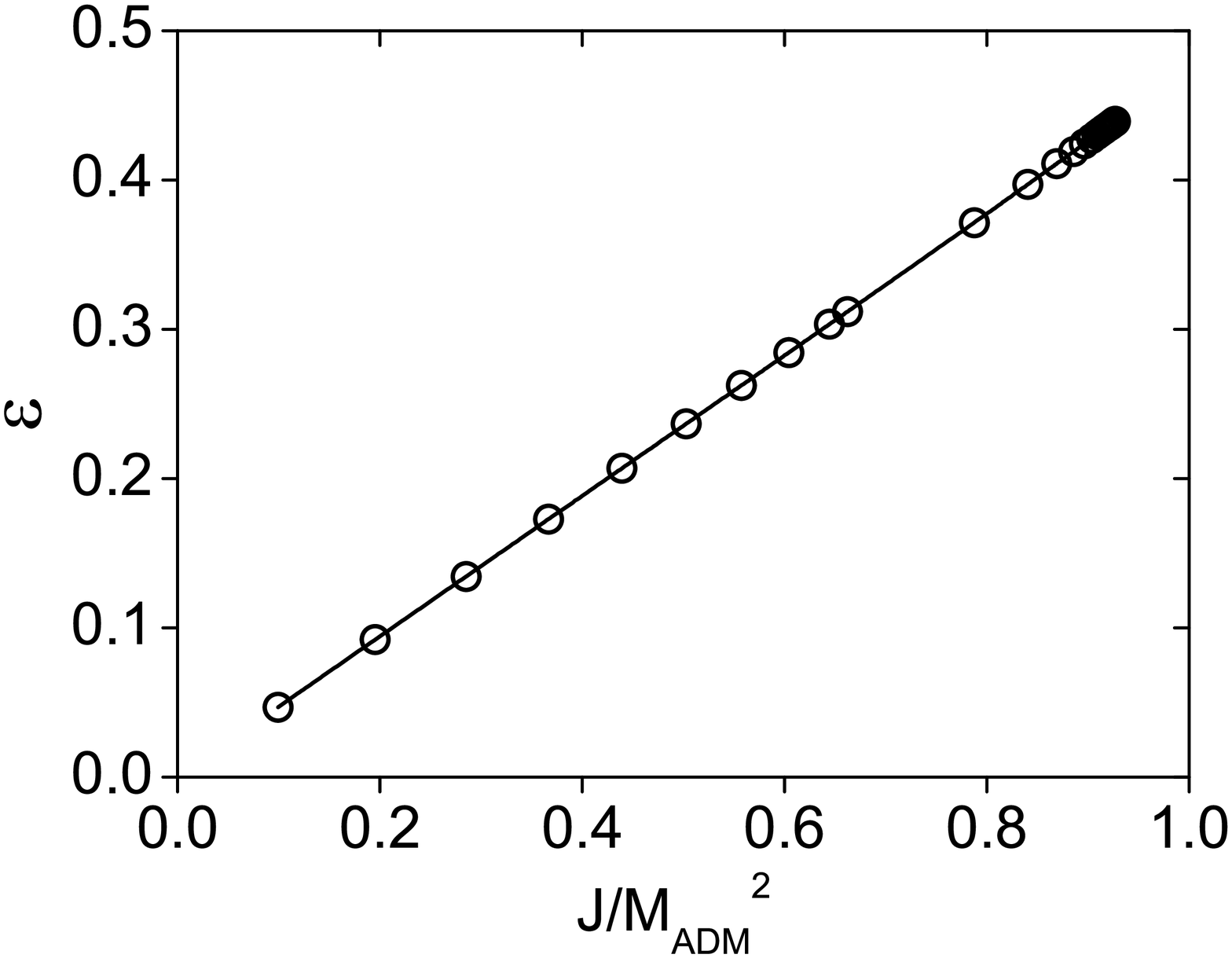}
\caption{Both graphs show the eccentricity of the minimal surface versus the $J_0/m_0^2$ and $J_0/M{ADM}^2$, respectively. The eccentricity tends to a limit value about $0.439$. In the inset the continuous line represents the approximate exact solution of Ref. \cite{immer_baumg} valid for small angular momentum parameter together with the numerical eccentricities.}
\end{figure}

We have revisited the estimate of the radiation content or the junk radiation present in the trumpet and wormhole initial data sets which have been considedred in Refs. \cite{hanann_id_trumpet,cook_phd,cook_york}. The radiation content, $E_{rad}$, is estimate as \cite{hanann_id_trumpet},

\begin{equation}
E_{rad} = \sqrt{M_{ADM}-P^2} - M_{BH}, \label{eq43}
\end{equation}

\noindent where $J^2=J_i J^i$, and $M_{irr}$ is the irreducible mass given by,

\begin{equation}
M_{BH} = M^2_{irr} + \frac{J^2}{4 M_{irr}^2}, \label{eq44}
\end{equation}

\noindent where $J^2=J_i J^i$, and the irreducible mass $M_{irr}$ is, 

\begin{equation}
M_{irr} = \sqrt{\frac{A}{16 \pi}}, \label{eq45}
\end{equation}

\noindent here $A$ is the area of the apparent horizon. After solving the apparent horizon equation for spinning and boosted black holes (see the Appendix B), $A$ can be calculated, allowing to determine the ratio $e_{rad} \equiv E_{rad}/M_{BH}$ in function of $j_0=J_0/M_{BH}^2$ and $p_0=P_0/M_{BH}$, respectively. We have noticed that for spinning black holes the radiation content in the trumpet and wormhole data is nearly the same. However, there is a slight exception for small $j_0$ in which $(e_{rad})_{\mathrm{trumpet}} > (e_{rad})_{\mathrm{wormhole}}$ (cf. Fig. 5). On the other hand, the amount of radiation in the trumpet and wormhole boosted black holes is indistinguishable according to the Fig. 5. 

\begin{figure}
\includegraphics[scale=0.3]{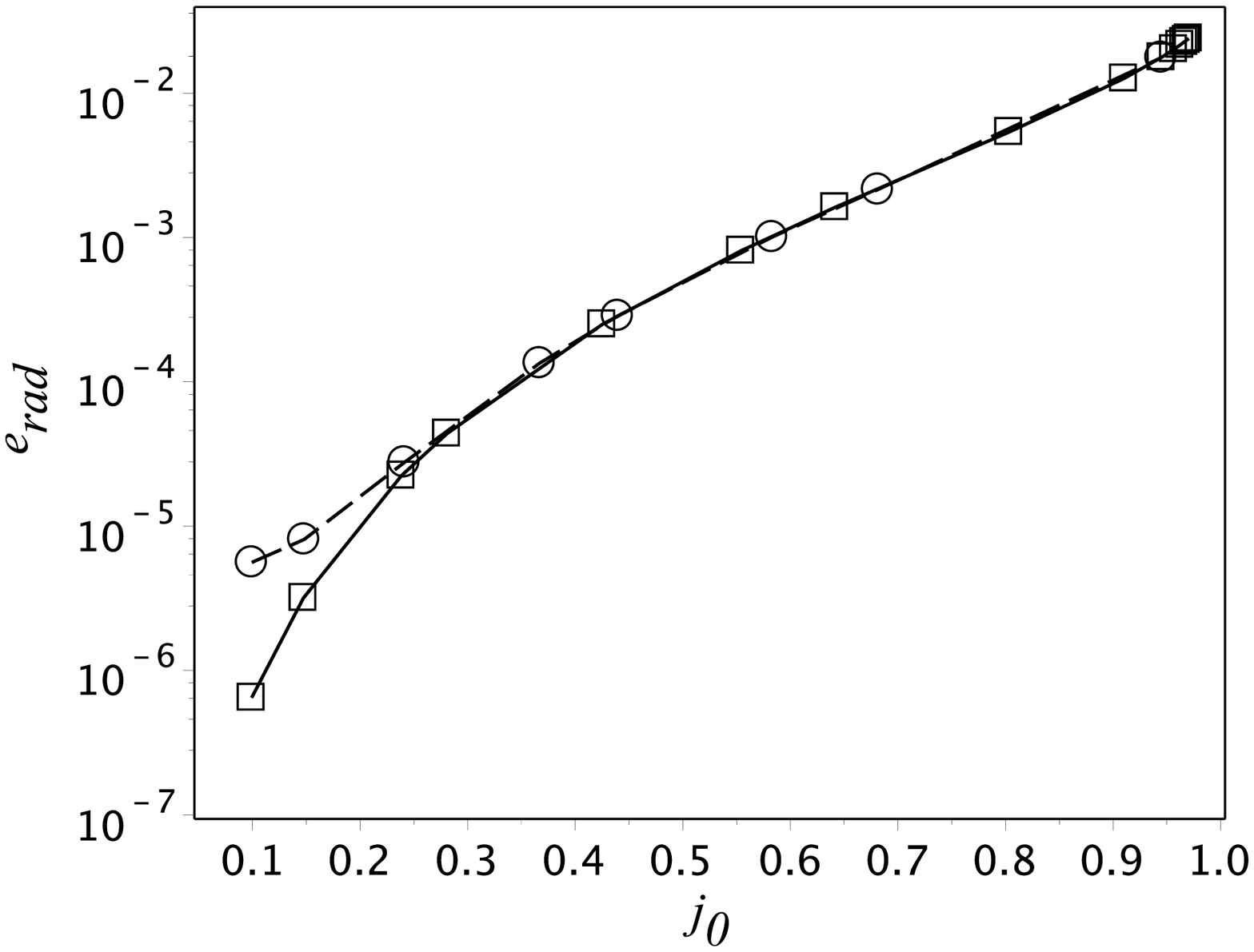}
\includegraphics[scale=0.3]{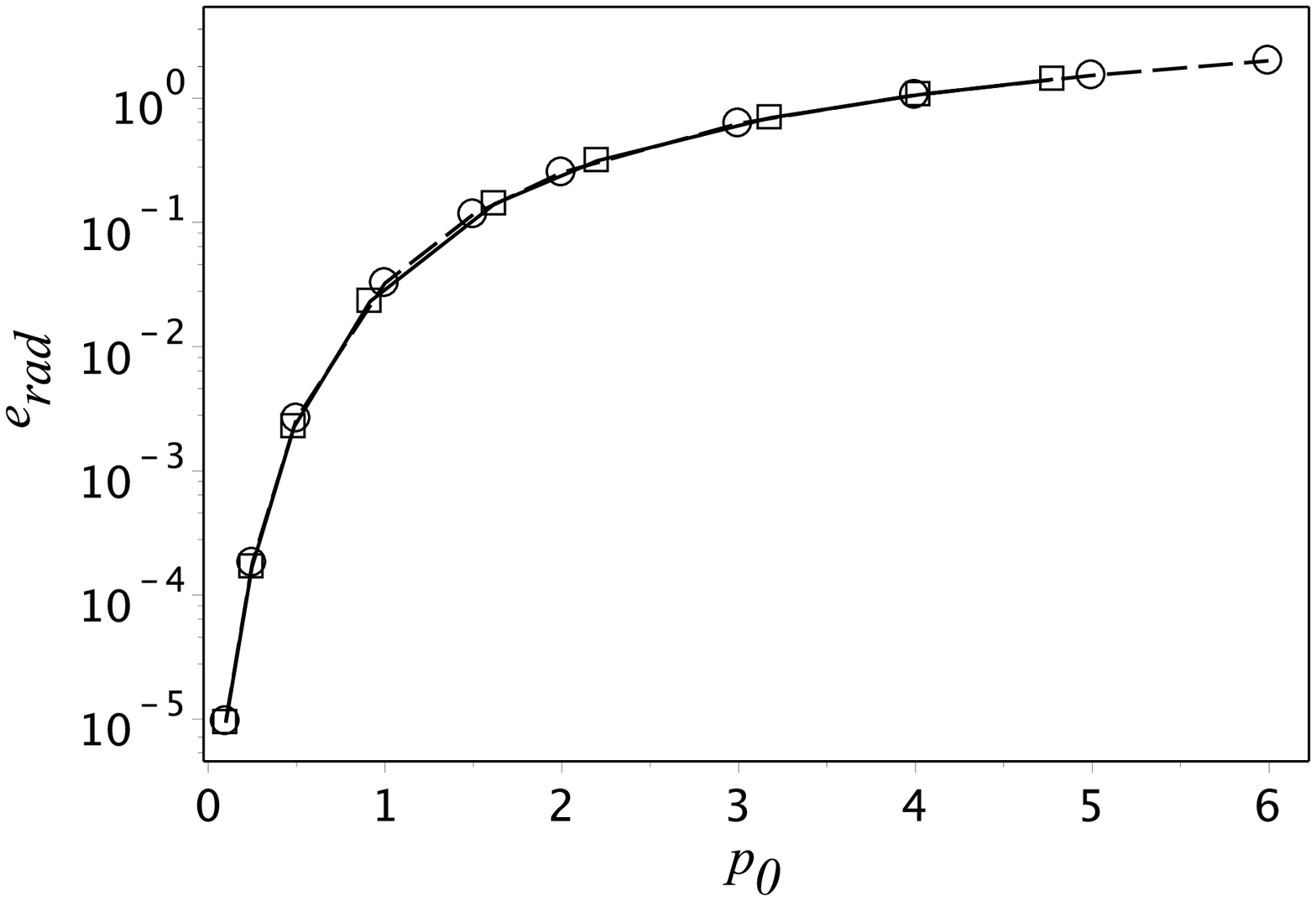}
\caption{Radiation content for spinning and boosted black holes (upper and down graphs). The circles and boxes refer to trumpet and wormhole data sets, respectively. In both cases $e_{rad} \equiv E_{rad}/M_{BH}$, and  $j_0=J_0/M_{BH}^2$ and $p_0=P_0/M_{BH}$.}
\end{figure}

To illustrate an application of the Galerkin-Collocation algorithm to a simple three-dimensional case, we have considered a trumpet puncture located at the origin and with linear and intrinsic angular momenta characterized, respectively by $\mathbf{P} = (P_0,0,0)$ and $\mathbf{S} = (0,0,J_0)$. In this case, 

\begin{eqnarray}
&&\bar{A}_{ij}\bar{A}^{ij} = \frac{18 J_0^2}{r^6} \sin^2\theta + \frac{9 P_0^2}{2r^4}(1+2 \sin^2\theta\,\cos^2\theta) + \nonumber \\
\\
&& + \frac{81 m_0^4}{8 r^6} - \frac{18 J_0 P_0}{r^5} \sin\theta \sin\phi - \frac{27 \sqrt{3}P_0 m_0^2}{2r^5} \sin\theta \cos\phi. \nonumber \label{eq46}
\end{eqnarray}

\noindent We have adopted the conformal factor as given by Eq. (\ref{eq15}) due to the presence of spin, and the relevant parameters are: $m_0=1$, $P_0=0.2 m_0$, and the spin parameter assumes several values, $J_0 = 0.1 m_0^2,0.2 m_0^2,..,0.5m_0^2$. The influence of increasing the spin parameter on the regular part of the conformal factor, $1+u(r,\theta,\phi)$, can be viewed in Fig. 6 showing the projection of $1+u$ on the plane $y=z=0$.	Notice the deformation produced by increasing $J_0$ by inspecting the curves from down to up.   

\begin{figure}
\includegraphics[height=5.5cm,width=6.0cm]{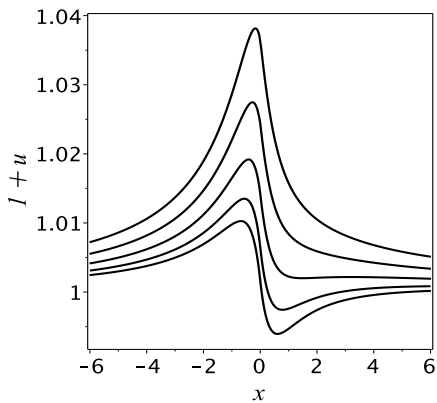}
\caption{Regular function $1+u(r,\theta,\phi)$ projected into the plane $y=z=0$ ($\theta=0,\phi=0,\pi$). We have fixed the boost parameter to $P_0=0.2 m_0$ while varying the spin as $J_0=0.1m_0^2,0.2m_0^2,..,0.5m_0^2$ with the corresponding profiles indicated by the curves from down to up.}
\end{figure}

\subsection{Binary black holes}

We discuss here a boosted binary formed with trumpet punctures lying at the axis z at the coordinate locations indicated by $\mathbf{C}_1=(0,0,-a)$ and $\mathbf{C}_2=(0,0,a)$ with $2a$ the coordinate separation between the punctures. We have adopted a simpler form of the conformal factor \cite{immer_baumg},

\begin{equation}
\Psi = \Psi_1 + \Psi_2 - 1 + u, \label{eq47}
\end{equation}

\noindent where $\Psi_1$ and $\Psi_2$ have the same form of $\Psi_0$ (see Appendix A) but are centered on $\mathbf{C}_1$ and $\mathbf{C}_2$ \cite{immer_baumg} respectively. Since the momentum constraint is a linear equation, the extrinsic curvature $\bar{A}_{ij}$ is given by,

\begin{equation}
\bar{A}_{ij} = \bar{A}_{ij}^{0(1)} + \bar{A}_{ij}^{0(2)} + \bar{A}_{ij}^{\mathbf{P}_1} + \bar{A}_{ij}^{\mathbf{P}_2}, \label{eq48}
\end{equation}

\noindent and the Hamiltonian constraint becomes,

\begin{eqnarray}
&&\bar{\nabla}^2 u + \frac{1}{8}(\Psi_1 + \Psi_2 - 1 + u)^{-7}\bar{A}^{ij}\bar{A}_{ij} \nonumber \\
\nonumber \\
&&- \frac{1}{8 \Psi_1^7}(\bar{A}_0^{ij}\bar{A}^0_{ij})^{(1)} - \frac{1}{8 \Psi_2^7}(\bar{A}_0^{ij}\bar{A}^0_{ij})^{(2)} = 0. \label{eq49}
\end{eqnarray}


\noindent The expression for $A_{ij}A^{ij}$ is shown in the Appendix C, and $(A^0_{ij}A_0^{ij})^{(1,2)}=81 m_{1,2}^2/r_{1,2}^6$. The algorithm presented in the last Section is straightforwardly adapted to solve the Hamiltonian constraint for trumpet binary punctures with the function $u$ approximated as indicated in Eq. (\ref{eq25}). The radial basis function is given by Eq. (\ref{eq29}). 

To test the algorithm, we have verified the convergence of the ADM mass for the axisymmetric binary system after setting $m_1=m_2=0.5$, $\mathbf{P}_1=(0,0,P_0)$, $\mathbf{P}_2=(0,0,-P_0)$, together with $a=3$ and $P_0=0.4 m_1$. Following the convergence test, we have fixed $N_y=14$, and the radial truncation order is made to vary as $N_x=20,25,30,..,100$.  Fig. 7 shows the exponential convergence of the ADM that is calculated according to Eq. (\ref{eq24})  in which $m_1+m_2$ replaces by $m_0$. In this case, the best choice for the map parameter is the coordinate separation between the punctures, $L_0=2a$. For the sake of illustration  we have included in Fig. 7 the plot of $1+u(r,\theta)$ in the plane $x=y=0$ for the binary black hole under consideration. 


As the last application, we have considered a three-dimensional binary formed by boosted wormhole punctures with $\mathbf{P}_1=(P_0,0,0)$ and $\mathbf{P}_2=(-P_0,0,0)$. The conformal factor is expressed in the same way as in Eq. (\ref{eq47}),

\begin{eqnarray}
\Psi = 1 + \frac{1}{2}\left(\frac{m_1}{r_1} + \frac{m_2}{r_2}\right) + u, \label{eq50}
\end{eqnarray}


\noindent Here the Hamiltonian constraint and the function $u$ are given respectively by Eqs. (\ref{eq18}) and (\ref{eq25}), and the corresponding expression for $A_{ij}A^{ij}$ is given in the Appendix C. The values of the parameters are the same of Ref. \cite{ansorg_1}: $a=3.0 M$, $m_1=m_2=0.5M$ and $P_0=0.2M$, where $M=m_1+m_2$. In Fig. 8 we show the two and the three dimensional plots of the $1+u(x,y=0,z)$. We have used truncation orders $N_x=40,N_y=16$ which means $40$ radial collocation points and a grid of $33 \times 33$ collocation points for the quadrature formulae given by expression (\ref{eq39}).

\begin{figure}
\includegraphics[height=5.cm,width=6cm]{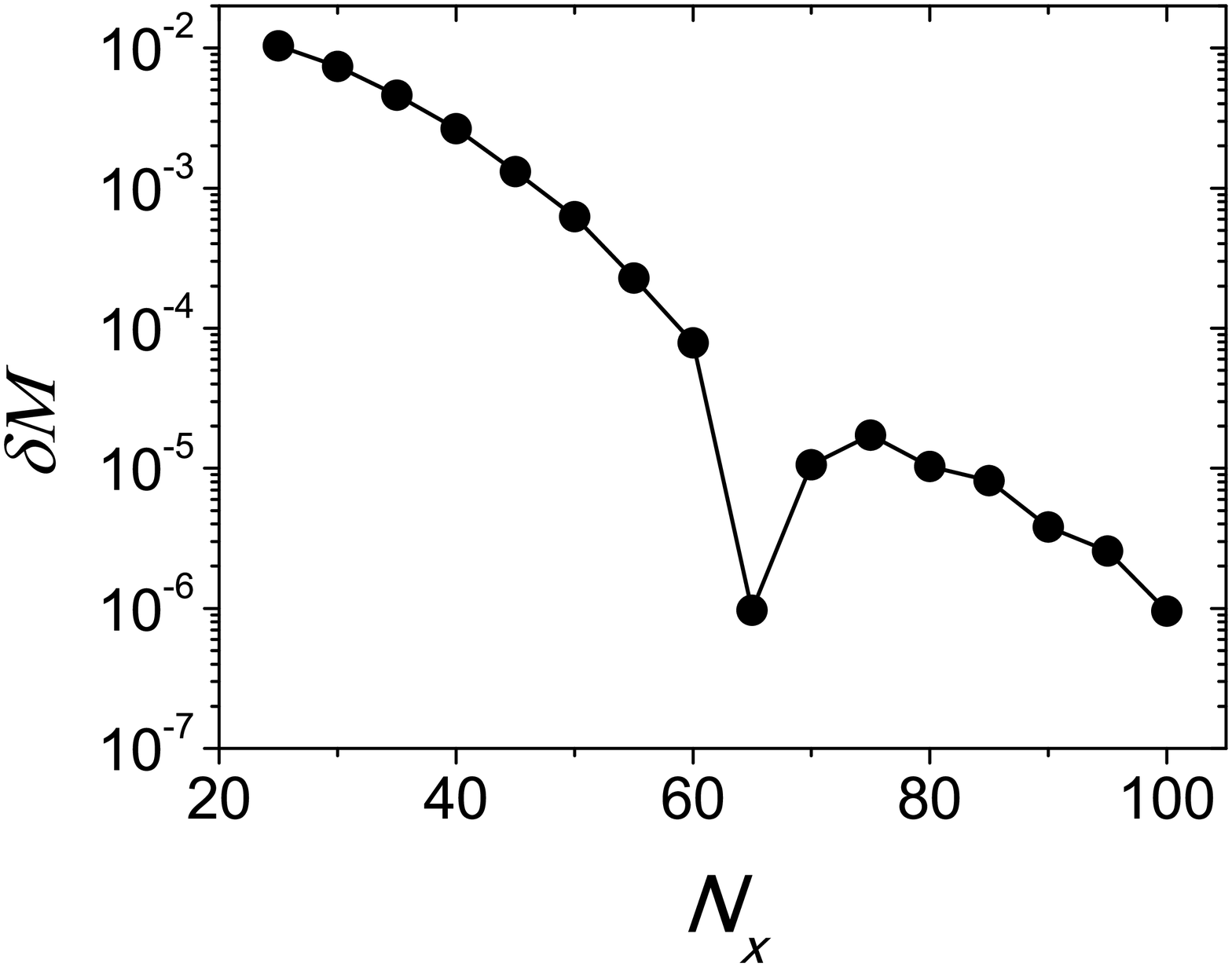}
\includegraphics[height=4.5cm,width=5cm]{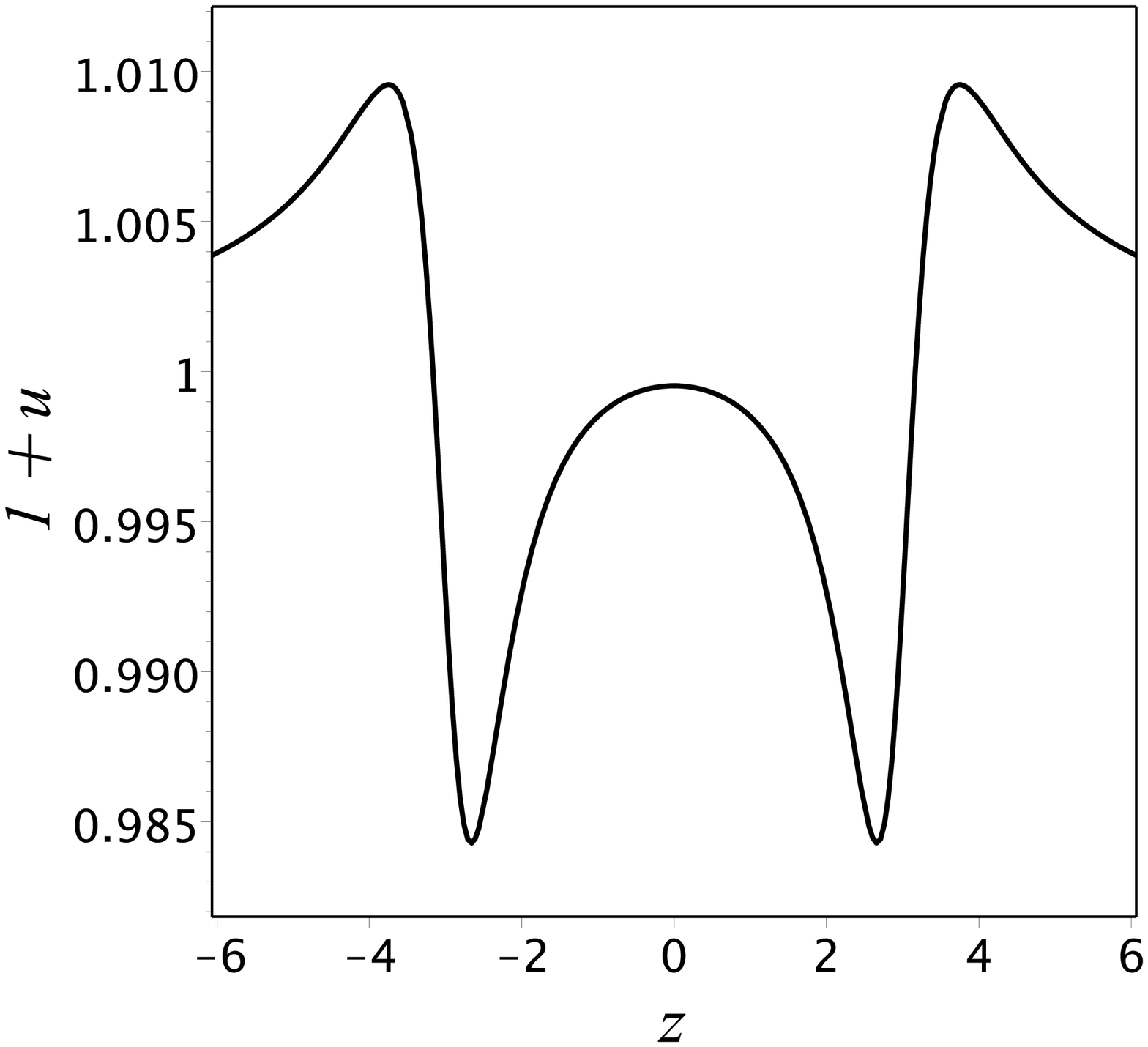}
\caption{Convergence of the ADM mass for the binary of boosted trumpet punctures located at the z-axis with $a=3$, $m_1=m_2=0.5$ and $P_0=0.4 m_1$. We have set $N_x=20,25,30,35,..,100$ and $N_y=14$. In the second panel we show the profile of $1+u$ ($N_x=100,N_y=14$) projected into the plane $x=y=0$.}
\end{figure}

\begin{figure}
\includegraphics[height=6.5cm,width=7.5cm]{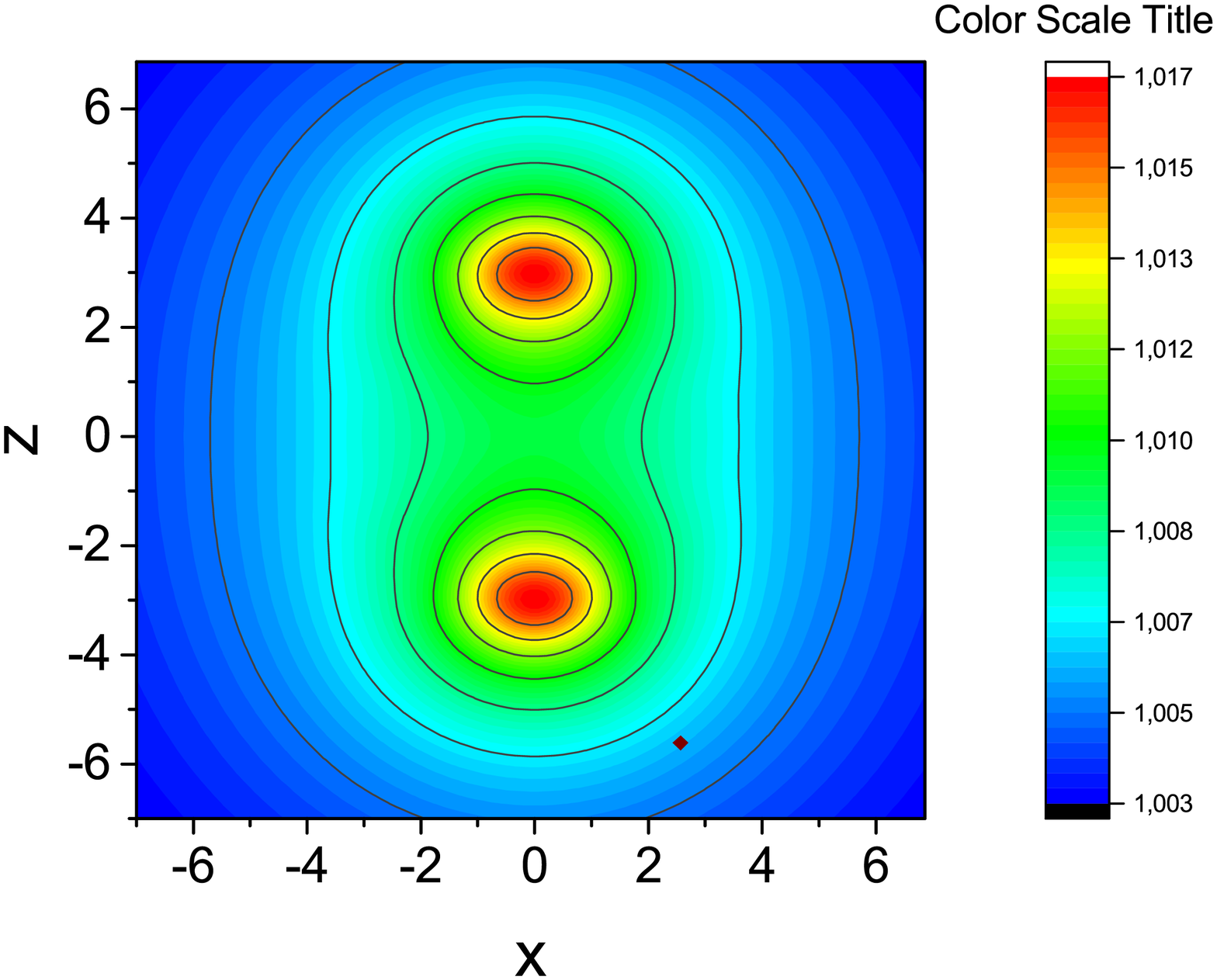}
\includegraphics[height=6.5cm,width=7.5cm]{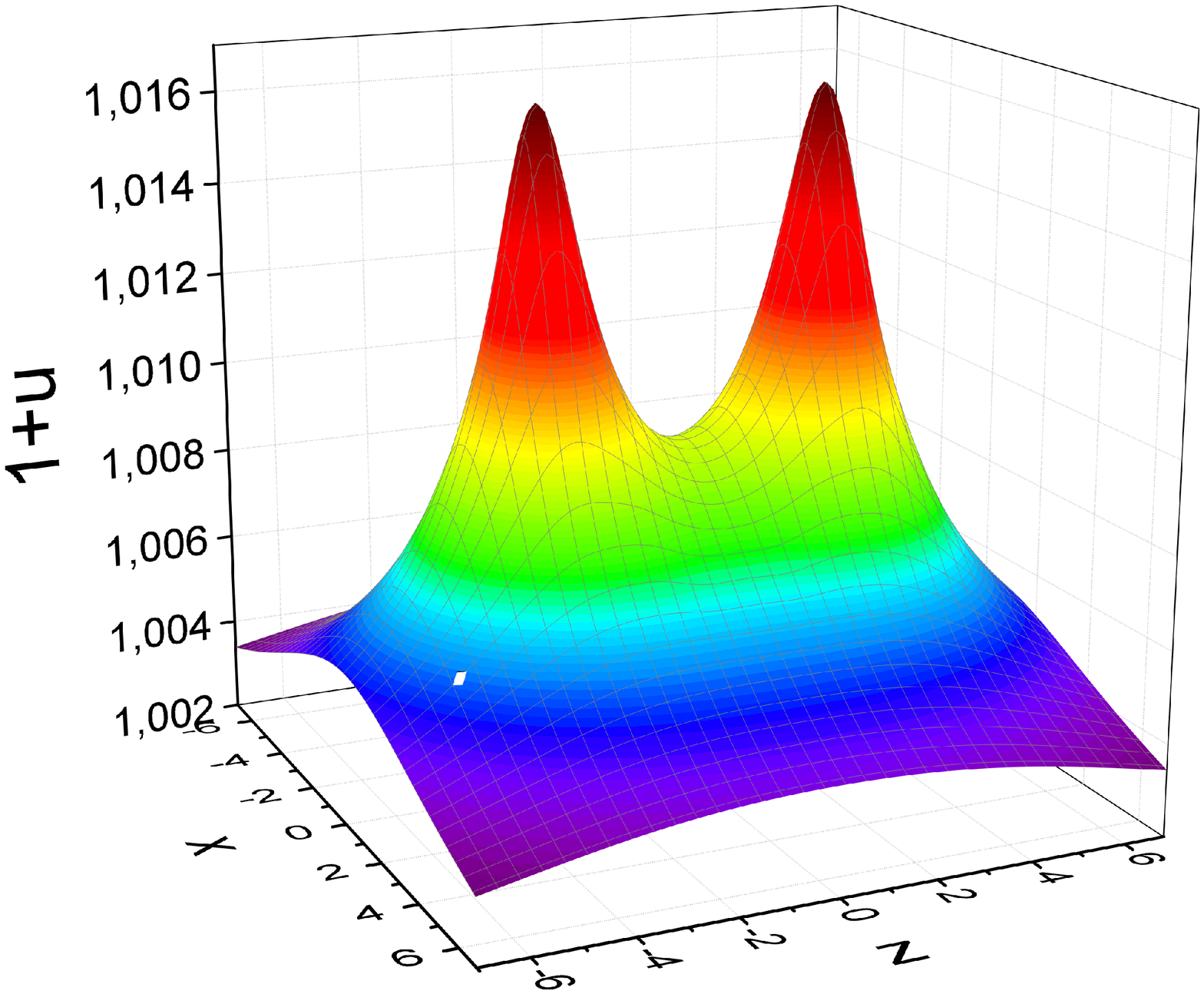}
\caption{Illustrations in two and three-dimensional plots (upper and lower panels, respectively) of $1+u$ for the binary of boosted wormhole punctures in which $a=3$, $m_1=m_2=0.5$ and $P_0=0.4 m_1$.}
\end{figure}

\section{Final remarks}%

We have presented a single domain algorithm using the Galerkin-Collocation method to solve the Hamiltonian constraint for trumpet and wormhole puncture data sets with emphasis for the first data sets.  We have considered Bowen-York data including the cases of spinning, boosted, single and binary black hole. We find worth of mentioning some features of the algorithm. The spatial domain is covered by spherical coordinates $(r,\theta,\phi)$. In all cases, the regular part of the conformal factor is approximated by Eq. (\ref{eq25}) with the radial basis functions satisfying the appropriate boundary conditions and taking the spherical harmonics as the angular basis functions. 

To describe trumped data corresponding to a single spinning and boosted black hole, we have proposed a puncture-like approach with a new form of the conformal factor given by expression (\ref{eq15}). We have also taken into account the analytical solution that describes the trumpet Schwarzschild black hole found by Baumgarte and Naculich \cite{baumg_nac} as the background solution. This procedure is analogous to make the explicit use of the background solution $\Psi_0=1+m_0/2r$ in the case of a single wormhole Schwarzschild black hole.

We have tested the algorithm successfully by checking the exponential convergence of the ADM mass that was present in most of the cases.  In the sequence, we have made some applications of the algorithm to situations of interest. Of particular importance is the case of a single spinning trumpet black hole, in which we have shown the influence of the spin in deforming the minimal surface from a spherical to an oblate spheroid by evaluating the eccentricity of the resulting surface. The eccentricity has the limit value of about $0.439$ obtained for large spin parameters. Interestingly, this value is approximate the half of the eccentricity of the ergosphere of the extremal Kerr black hole.

We have revisited the amount of radiation content present in the trumpet and wormhole single spinning and boosted black holes. In general, the radiation content is nearly the same in both families of initial data sets as indicated by Fig. 5. We have also presented the profiles of the regular function $u(r,\theta,\phi)$ for the single trumpet black hole with spin and boost. By fixing the boost parameter $P_0$ and decreasing the spin $J_0$ we noticed that the profile approach to that corresponding to single boosted black hole as expected. 

For the last and more illustrative applications of the algorithm, we have considered initial data for trumpet and wormhole binaries. Trumpet data constituted by binary boosted black holes was envisaged for the axisymmetric case; the ADM mass converges exponentially. For a more general case, we generate an initial data with wormhole boosted black holes with the same parameters of Ref. \cite{ansorg_1} but with truncation orders $N_x=40$ and $N_{\theta}=16$, which means 40 radial collocation points, and a grid of $33 \times 33$ angular points for the quadrature formulae (\ref{eq39}). 

The Galerkin-Collocation method is a viable alternative to solve the Hamiltonian constraint for the trumpet and wormhole initial data sets. We point out two directions to follow. The first is to consider $1+\mathrm{log}$ trumpet data sets for which the maximal sliced conditions is relaxed \cite{hannan_mov_punct3,dietrich_brugmann}. The second is to extend the present algorithm including more than one domain using the technique of domain decomposition.

\section*{Acknowledgements}

The authors acknowledge the financial support of the Brazilian agencies CNPq, CAPES and FAPERJ. HPO thanks FAPERJ for support within the grant BBP (Bolsas de Bancada para Projetos). We also would like to thank Thomas W. Baumgarte for comments on the manuscript.


\appendix


\section{Background Schwarzschild trumpet exact solution}

The exact expression corresponding to the maximally sliced trumpet of the Schwarzschild spacetime was derived by Baumgarte and Naculich \cite{baumg_nac}:
{\small
\begin{eqnarray}
\Psi_0 &&= \left[\frac{4 R}{2R+m_0+\sqrt{4R^2+4m_0R+3m_0^2}}\right]^{1/2} \\
\nonumber \\
&& \times \left[\frac{8R+6m_0+3\sqrt{8R^2+8m_0R+6m_0^2}}{(4+3\sqrt{2})(2R-3m_0)}\right]^{1/2\sqrt{2}} \nonumber
\end{eqnarray}}

\noindent where the isotropic radial coordinate $r$ is,
 
{\small 
\begin{eqnarray}
r &&= \left[\frac{2R+m_0+\sqrt{4R^2+4m_0R+3m_0^2}}{4}\right] \\
\nonumber \\
&& \times \left[\frac{(4+3\sqrt{2})(2R-3m_0)}{8R+6m_0+3\sqrt{8R^2+8m_0R+6m_0^2}}\right]^{1/\sqrt{2}} \nonumber
\end{eqnarray}}

We have located the binary punctures along the z-axis ($\mathbf{C}_{1,2}=(0,0,\pm a)$) for the sake of convenience. The background conformal factors have the same form of Eq. (A1), however with $\Psi_1=\Psi_1(R_1)$ and $\Psi_2=\Psi_2(R_2)$. The relation between the areal radius $R_1$ with the coordinates $(r,\theta)$ is

{\small 
\begin{eqnarray}
&&\sqrt{r^2+2ar\cos\theta+a^2} = \left[\frac{2R_1+m_1+\sqrt{4R_1^2+4m_1R_1+3m_1^2}}{4}\right] \nonumber \\
\nonumber \\
&& \times \left[\frac{(4+3\sqrt{2})(2R_1-3m_1)}{8R_1+6m_1+3\sqrt{8R_1^2+8m_1R_1+6m_1^2}}\right]^{1/\sqrt{2}},
\end{eqnarray}}

\noindent and a similar expression connecting $R_2$ with $(r,\theta)$.

\section{The apparent horizon}

The apparent horizon for axisymmetric systems satisfies the following ordinary differential equation,

{\small
\begin{eqnarray}
\partial^2_{\theta} h &=& -\Gamma_{BC}^A M_A u^B u^C - \left(\frac{ds}{d\theta}\right)^2 \gamma^{\phi\phi} \Gamma^A_{\phi\phi} m_A - \left(\gamma^{(2)}\right)^{-1/2}\nonumber \\
&&  \times \frac{ds}{d\theta} u^A u^B K_{AB} - \left(\gamma^{(2)}\right)^{-1/2} \left(\frac{ds}{d\theta}\right)^3 \gamma^{\phi\phi} K_{\phi\phi},
\end{eqnarray}}

\noindent where $r=h(\theta)$ describes the apparent horizon surface, $m_i=(1,-\partial_\theta h,0)$, $u^i=(\partial_\theta h,1,0)$ and $(ds/d\theta)^2 = \gamma_{AB} u^A u^B$; the capital letters run over the coordinates $r,\theta$. Since $K=0$ it follows that $K_{ij} = A_{ij} = \Psi^2 \bar{A}_{ij}$. The conformal factor is obtained after solving numerically the Hamiltonian constraint and inserted into the apparent horizon equation.

We have introduced $\tilde{y}=\cos\theta$ and transformed the apparent horizon equation in a non-autonomous dynamical system of the type $\partial_{\tilde{y}} h = v, \partial_{\tilde{y}} v = f(h,v,\tilde{y})$ whose solution must satisfy the boundary conditions $\partial_\theta h=0$ for $\theta=0,\pi$ or $v \sqrt{1-\tilde{y}} = 0$ for $\tilde{y}=-1,1$.

\begin{figure}[ht]
\includegraphics[height=6.5cm,width=6.5cm]{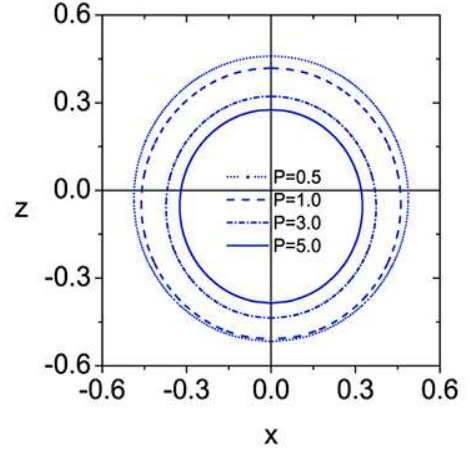}
\caption{Locations of the apparent horizon $r=h(\theta)$ for a boosted black hole in the wormhole representation for several values of the momentum parameter along the z-axis. The corresponding locations for the trumpet representation are similar.}
\end{figure}

\section{Extrinsic curvature for binary black holes}

The quantity $\bar{A}_{ij}\bar{A}^{ij}$ for trumpet boosted punctures with $\mathbf{P}_1=(0,0,P_1), \mathbf{P}_2=(0,0,P_2)$ and located at $\mathbf{C}_1=(0,0,-a),\mathbf{C}_2=(0,0,a)$, respectively, is given by,

{\small 
\begin{widetext}
\begin{eqnarray}
\bar{A}_{ij}\bar{A}^{ij} &&= \frac{9 P_1^2}{2 r_1^6} \left[(1+2\cos^2\theta)r^2 + 6ar\cos\theta + 3a^2\right] + \frac{9 P_2^2}{2 r_2^6} \left[(1+2\cos^2\theta)r^2 - 6ar\cos\theta + 3a^2\right] + \frac{9 P_1 P_2}{2 r_1^5 r_2^5} \big[(1+2\cos^2\theta)r^6 \nonumber \\
&&+ (2\cos^4\theta-14\cos^2\theta+3)a^2r^4 +(8\cos^2\theta+1)a^4r^2 - 3a^6\big] + \frac{81m_1^4}{8 r_1^6} + \frac{81m_2^4}{8 r_2^6} + \frac{81 m_1^2 m_2^2}{4r_1^5r_2^5}(2a^2r^2\cos^2\theta+a^4-4a^2r^2 \nonumber \\
&&+r^4) -\frac{27\sqrt{3}m_1^2 P_1}{2r_1^6}(r\cos\theta+a) -\frac{27\sqrt{3}m_2^2 P_2}{2r_2^6}(r\cos\theta-a) -\frac{27\sqrt{3}m_2^2P_1}{2r_1^5r_2^5}\big[a^5+a^4r\cos\theta-2a^3r^2+(2\cos^2\theta-4)\cos\theta a^2r^3 \nonumber \\
&&+(2\cos^2\theta-1)ar^4+r^5\cos\theta \big] + \frac{27\sqrt{3}m_1^2 P_2}{2r_1^5 r_2^5}\big[a^5-a^4r\cos\theta-2a^3r^2+(-2\cos^2\theta+4)a^2r^3\cos\theta+(2\cos^2\theta-1)ar^4-r^5\cos\theta\big] \nonumber \\
\end{eqnarray}
\end{widetext}}

\noindent where $r_1=\sqrt{r^2+2ar\cos\theta+a^2}$ and $r_1=\sqrt{r^2-2ar\cos\theta+a^2}$. For the case of wormhole boosted punctures located at $\mathbf{C}_1,\mathbf{C}_2$ with $\mathbf{P}_1=(P_1,0,0), \mathbf{P}_2=(P_2,0,0)$, we have, 

{\small
\begin{widetext}
\begin{eqnarray}
\bar{A}_{ij}\bar{A}^{ij} &=& \frac{9 P_1^2}{2 r_1^6}\big[2ar\cos\theta+a^2+r^2+2r^2(1-\cos^2\theta)\cos^2 \phi\big] + \frac{9 P_2^2}{2 r_2^2}(-2ar\cos\theta+a^2+r^2+2r^2(1-\cos^2\theta)\cos^2\phi)+\frac{9 P_1 P_2}{2 r_1^3 r_2^3} \nonumber \\
&& \times \left[r^2-a^2+\frac{2r^2(1-\cos^2\theta)(r^4-a^4-a^2r^2(1-\cos^2\theta))}{(2ar\cos\theta+a^2+r^2)(r^2-2ar\cos\theta+a^2)}\cos^2\phi\right]
\end{eqnarray}
\end{widetext}
}


\begin{thebibliography}{99}

\bibitem{cook} G. B. Cook, Initial Data for Numerical Relativity, Liv. Rev. Relativity, 3, 5 (2000).

\bibitem{LIGO_gws} B. P. Abbott et al. (LIGO Scientific Collaboration and Virgo Collaboration), Phys. Rev. Lett. 116, 061102 (2016).

\bibitem{pretorius} F. Pretorius, Phys. Rev. Lett. \textbf{95}, 121101 (2005).

\bibitem{campanelli} M. Campanelli, C. O. Lousto, P. Marroneti and Y. Zlochower, Phys. Rev. Lett. \textbf{96}, 111102 (2006).

\bibitem{baker} J. G. Baker, J. Centrella, D. -I. Choi, M. Koppitz and J. van Meter, Phys. Rev. Lett. \textbf{96}, 111102 (2006).

\bibitem{adm} R. Arnowitt, S. Deser and C. W. Misner, in \textit{Gravitation: an Introduction to Current Research}, edited by L. Witten (Willwy, 1962), p.227.

\bibitem{york_1979}J. W. York, Jr., in \textit{Sources of Gravitational Radiation}, edited by L. L. Smarr (Cambridge University Press, London, 1979), p. 83.

\bibitem{bowen_york} J. M. Bowen and J. W. York, Phys. Rev. D \textbf{21}, 2047 (1980).

\bibitem{baumgarte_shapiro} Thomas W. Baumgarte and Stuart L. Shapiro, \textit{Numerical Relativity, Solving the Einstein's Equations on the Computer}, Cambridge University Press (2010).

\bibitem{brandt_brugmann} S. Brandt adn B. Brugmann, Phys. Rev. Lett. \textbf{78}, 3606 (1997).

\bibitem{brugmann_04} B. Brugmann, W. Tichy and N. Jansen, Phys. Rev. Lett. \textbf{92}, 211101 (2004).

\bibitem{baumg_2000} T. W. Baumgarte, Phys. Rev. D \textbf{62}, 024018 (2000).

\bibitem{diener_06} P. Diener, F. Herrman, D. Pollney, E. Schnetter, E. Seidel, R. Takahashi, J. Thornburg and J. Centrella, Phys. Rev. Lett. \textbf{96}, 121101 (2006).

\bibitem{baker_06} J. G. Baker, J. Centrella, Dae-Il Choi, M. Koppitz and J. R. van Meter, Phys. Rev. D \textbf{73}, 104002 (2006).

\bibitem{meter_06} J. R. van Meter, J. G. Baker, M. Koppitz and Dae-Il Choi, Phys. Rev. D \textbf{73}, 124011 (2006).

\bibitem{bode_09} T. Bode, P. Laguna, D. M. Schoemaker, I. Hinder, F. Hermann and B. Vaishnav, Phys. Rev. D \textbf{80}, 024008 (2009).

\bibitem{hannan_mov_punct} M. Hannan, S. Husa, N. O. Murchadha, B. Brugmann, J. A. Gonzalez and U. Sperharke, J. Phys. Conf. Series \textbf{66}, 01247 (2007).

\bibitem{hannan_mov_punct2} M. Hannan, S. Husa, D. Pollney, B. Brugmann and N. O. Murchadha, Phys. Rev. Lett. \textbf{99}, 241102 (2007).

\bibitem{brown} J. D. Brown, Phys. Rev. D \textbf{77}, 044018 (2008).

\bibitem{hannan_mov_punct3} M. Hannan, S. Husa, F. Ohme, B. Brugmann and N. O. Murchadha, Phys. Rev. D \textbf{78}, 064020 (2008).

\bibitem{baumg_nac} T. W. Baumgarte and S. G. Naculich, Phys. Rev. D \textbf{75}, 067502 (2007).

\bibitem{denninson_baumg} K. A. Denninson and T. W. Baumgarte, Class. Quantum Grav. \textbf{31}, 117001 (2014).

\bibitem{hanann_id_trumpet} M. Hanann, S. Husa and N. O. Murchadha, Phys. Rev. D \textbf{80}, 124007 (2009).

\bibitem{immer_baumg} Jason D. Immerman, T. W. Baumgarte, Phys. Rev. D \textbf{80}, 061501(R) (2009).

\bibitem{boyd} John P. Boyd, \textit{Chebyshev and Fourier Spectral Methods}, Dover Publications (2001).

\bibitem{dietrich_brugmann} T. Dietrich and B. Brugmann, Phys. rev. D \textbf{89}, 024014 (2014).

\bibitem{deol_rod_bondi} H. P. de Oliveira and E. L. Rodrigues, Class. Quant. Grav, \textbf{28}, 235011 (2011).

\bibitem{deol_rod_RT} H. P. de Oliveira, E. L. Rodrigues and J. F. E. Skea, Phys. Rev. D \textbf{84}, 044007 (2011).


\bibitem{deol_rod_idata2} H. P. de Oliveira and E. L. Rodrigues, Phys. Rev. D \textbf{86}, 064007 (2011).

\bibitem{pfeiffer_CPC} Harald P. Pfeiffer, Lawrence E. Kidder, Mark A. Scheel and Saul Teukolsky, Comp. Phys. Commun. \textbf{152}, 253 (2003).

\bibitem{ansorg_1} Marcus Ansorg, Bernd Brugmann and Wolfang Tichy, Phys. Rev. D \textbf{70}, 064011 (2004).

\bibitem{ansorg_07} Marcus Ansorg, Class. Quant. Grav. \textbf{24}, S1-14 (2007).

\bibitem{ossokine} S. Ossokine, F. Foucart, H. P. Pfeiffer, M. Boyle and B. Szilagyi, Class. Quant. Grav. \textbf{32}, 245010 (2015).















\bibitem{finlayson} B. A. Finlayson, \textit{The Method of Weighted Residuals and Variational Principles} (Academic Press, New York, 1972).

\bibitem{fornberg} B. Fornberg, \textit{A Pratical Guide to Pseudospectral Methods}, Cambridge Monographs on Applied and Computational Mathematics, Cambrige University Press (1998).

\bibitem{cook_phd} G. B. Cook, Ph.D. thesis, University of North Carolina at Chappel Hill, Chappel Hill, North Carolina (1990).

\bibitem{cook_york} G. B. Cook and J. W. York, Phys. Rev. D \textbf{41}, 1077 (1990). 

\end{thebibliography}
\end{document}